\documentclass{aa}
\usepackage{graphicx}
\begin{document}
\title{Supernova remnant S\,147 and its associated neutron star(s)}

\author{V.V.Gvaramadze\inst{1,2,3}\thanks{{\it Address for
correspondence}: Krasin str. 19, ap. 81, Moscow 123056, Russia
(vgvaram@sai.msu.ru)}}

\institute{Abdus Salam International Centre for Theoretical Physics,
Strada Costiera 11, PO Box 586, 34100 Trieste, Italy \and Sternberg
Astronomical Institute, Moscow State University, Universitetskij Pr.
13, Moscow 119992, Russia \and Center for Plasma Astrohysics,
E.K.Kharadze Abastumani Astrophysical Observatory, A.Kazbegi ave.
2-a, Tbilisi 0160, Georgia}

\date{Received / Accepted }

\titlerunning{Supernova remnant S\,147 and its associated neutron star(s)}
\authorrunning{Gvaramadze}

\abstract{The supernova remnant S\,147 harbors the pulsar PSR
J\,0538+2817 whose characteristic age is more than an order of
magnitude greater than the kinematic age of the system (inferred
from the angular offset of the pulsar from the geometric center of
the supernova remnant and the pulsar proper motion). To reconcile
this discrepancy we propose that PSR J\,0538+2817 could be the
stellar remnant of the first supernova explosion in a massive
binary system and therefore could be as old as its characteristic
age. Our proposal implies that S\,147 is the diffuse remnant of
the second supernova explosion (that disrupted the binary system)
and that a much younger second neutron star (not necessarily
manifesting itself as a radio pulsar) should be associated with
S\,147. We use the existing observational data on the system to
suggest that the progenitor of the supernova that formed S\,147
was a Wolf-Rayet star (so that the supernova explosion occurred
within a wind bubble surrounded by a massive shell) and to
constrain the parameters of the binary system. We also restrict
the magnitude and direction of the kick velocity received by the
young neutron star at birth and find that the kick vector should
not strongly deviate from the orbital plane of the binary system.
           }

\maketitle

\section{Introduction}
%
It is generally accepted that the supernova remnant (SNR)
\object{S\,147} is associated with the radio pulsar \object{PSR
J\,0538+2817}. The only solid argument in support of this
association is the positional coincidence of both objects: the
pulsar is located (at least in projection) well within the
extended shell of the SNR (Anderson et al. \cite{and96}). The
numerous estimates of the distance to the SNR are, in general, not
inconsistent with the dispersion measure distance to the pulsar.
The basic problem for the association is the obvious discrepancy
(Kramer et al. \cite{kra03}) between the kinematic age of the
system of $\sim 3\times 10^4$ yr (estimated from the angular
offset of the pulsar from the geometric center of the SNR and the
pulsar proper motion) and the characteristic (spin-down) age of
the pulsar of $\sim 6\times 10^5$ yr. To reconcile these ages one
can assume that the pulsar was born with a spin period close to
the present one (Kramer et al. \cite{kra03}; see also Romani \& Ng
\cite{rom03}). This assumption is often exploited to explain the
similar age discrepancy inherent to several other neutron star
(NS)/SNR associations (e.g. Migliazzo et al. \cite{mig02}).

In this paper, we propose an alternative explanation of the age
discrepancy because PSR J\,0538+2817 could be the stellar remnant
of the first supernova (SN) explosion in a massive binary system
and therefore could be as old as indicated by its characteristic
age (cf. Morris et al. \cite{mor78}). Our proposal implies that
S\,147 is the diffuse remnant of the second SN explosion (that
disrupted the binary system) and that a much younger second NS
(not necessarily manifesting itself as a radio pulsar) should be
associated with S\,147. In Sect.\,2 we review the existing
observational data on the system PSR J\,0538+2817/SNR S\,147. In
Sect.\,3 we suggest that the progenitor of the SN that formed
S\,147 was a Wolf-Rayet (WR) star and that the SN explosion
occurred within a wind bubble surrounded by a massive shell. In
Sect.\,4 we consider the possibility that PSR J\,0538+2817 is the
remnant of the first SN explosion in a massive binary. Sect.\,5
deals with some issues related to the content of the paper.

\section{The SNR S\,147 and PSR J\,0538+2817: observational data}

\subsection{General structure of S\,147}

S\,147 (also \object{G\,180.0$-$1.7}, \object{Simeis\,147},
\object{Shajn\,147}, etc) is a shell-type SNR with a diameter of
$\sim 3\degr$. The optical image of S\,147 presented by van den
Bergh et al. (\cite{van73}) shows a filamentary shell with a sharp
circular boundary to the south. The north boundary of the shell is
less regular: it consists of a long arc stretched in the east-west
direction and two lobes protruding beyond the arc in the northeast
and northwest directions for about one third of the characteristic
radius of the SNR. There are some indications that the northern
half of S\,147 expands somewhat faster than the southern one (see
Fig.\,2 of Lozinskaya \cite{loz76}). The east and west edges of
the SNR show signatures of blow-ups (more obvious to the east),
that makes the SNR somewhat elongated in the east--west direction.
The blow-ups are seen more prominently in the recent excellent
${\rm H}_{\alpha}$ image of S\,147 presented by Drew et al.
(\cite{dre05}; see Fig.~\ref{S147})\footnote{For other recent
images of S\,147 see
http://www.skyfactory.org/simeis147/simeis147.htm}. The 1.6 GHz
image of S\,147 by Kundu et al. (\cite{kun80}) shows a break
between the south and north halves of the SNR, typical of SNRs
with bilateral symmetry. The bilateral axis of S\,147, defined by
the radio brightness distribution, is parallel to the east--west
elongation of the SNR's shell.

Optical and ultraviolet observations of S\,147 (e.g. Lozinskaya
\cite{loz76}; Kirshner \& Arnold \cite{kir79}; Phillips et al.
\cite{phi81}) suggest that the expansion velocity of its shell is
$\simeq 80-120 \, {\rm km}\,{\rm s}^{-1}$, that implies that the
SNR has already entered the final (momentum-conserving) stage of
evolution (see, however, Sect.\,3). The same conclusion can be
derived from the good positional agreement between several optical
and radio filaments (Sofue et al. \cite{sof80}; see also F\"{u}rst
\& Reich \cite{fur86}) and from the non-detection of X-ray
emission from the SNR's shell (Souvageot et al. \cite{sou90}).
\begin{figure}
\centering
 \caption{The ${\rm H}_{\alpha}$ image of the supernova remnant S\,147
(Drew et al. \cite{dre05}; reproduced with permission of the IPHAS
collaboration). Position of the pulsar PSR J\,0538+2817 is
indicated by a cross. The line drawn in the east-west direction
shows the bilateral symmetry axis (see text for details). North is
up, east at left.}
  \label{S147}
\end{figure}

\subsection{PSR J\,0538+2817}

PSR J\,0538+2817 was discovered in the untargeted pulsar survey
conducted with the Arecibo radio telescope (Anderson et al.
\cite{and96}), which covered a $1\degr$ strip in the north half of
S\,147. The pulsar is located $\simeq 40\arcmin$ northwest of the
geometric center of the SNR. The spin period $P\simeq 0.143$ s and
the period derivative $\dot{P} \simeq 3.67\times 10^{-15} \, {\rm
s} \, {\rm s}^{-1}$ of the pulsar (Anderson et al. \cite{and96})
yield the characteristic age $\tau =P/2\dot{P} \simeq 6.2\times
10^5$ yr, the magnetic field strength $B = 3.2\times 10^{19} \,
(P\dot{P})^{1/2} \simeq 7.3\times 10^{11}$ G and the spin-down
luminosity $|\dot{E}| = 4\pi ^2 IP^{-3} \dot{P} \simeq 4.9\times
10^{34} \, {\rm erg} \, {\rm s}^{-1}$, where $I\simeq 10^{45} \,
{\rm g} \, {\rm cm}^{-2}$ is the moment of inertia of the pulsar.

The {\it Chandra X-ray Observatory} revealed a compact nebula
surrounding PSR J\,0538+2817 (Romani \& Ng \cite{rom03}). Romani
\& Ng (\cite{rom03}) interpreted the nebula as an equatorial torus
produced by the relativistic pulsar wind, similar to those
observed around several other rotation-powered pulsars located in
the high pressure interiors of their associated SNRs (e.g. Ng \&
Romani \cite{ng04}). Ng \& Romani (\cite{ng04}) found that the
symmetry axes of the toroidal nebulae (it is believed that they
coincide with the pulsar spin axes) show a trend (most prominent
in the case of the Crab and Vela pulsars) toward alignment with
the proper motion vectors of the pulsars. Based on this trend,
Romani \& Ng (\cite{rom03}) predicted the direction of the proper
motion of PSR J\,0538+2817 at a position angle of $\simeq
334\degr$ (measured north through east), which is consistent with
the direction from the geometric center of S\,147 to the present
position of the pulsar (see, however, Sect.\,5).

Kramer et al. (\cite{kra03}) used timing observations of PSR
J\,0538+2817 to measure its proper motion in the ecliptic
coordinates, $\mu _{\lambda} =-41 \pm 3 \, {\rm mas} \, {\rm
yr}^{-1}$ and $\mu _{\beta} =47 \pm 57 \, {\rm mas} \, {\rm
yr}^{-1}$, with a median value $\mu =67 _{-22} ^{+48} \, {\rm mas}
\, {\rm yr}^{-1}$ and a position angle of $311\degr _{-56\degr}
^{+28\degr}$. The position angle is consistent with the movement
of PSR J\,0538+2817 away from the geometric center of S\,147.
Kramer et al. (\cite{kra03}) consider this result as strong
evidence that the pulsar is associated with the SNR (cf.
Sect.\,2.3). The large error in the proper motion measurement in
the latitudinal direction, however, allows the possibility that
the pulsar trajectory is significantly offset (up to $\simeq
0.5\degr$) from the geometric center of S\,147.

\subsection{Age of the system PSR J\,0538+2817/SNR S\,147}

For a long time it was believed that S\,147 is one of the oldest
evolved SNRs in the Galaxy. This belief was based on the age
estimates of $1-2\times 10^5$ yr derived with help of the
Sedov-Taylor solution or relationships for radiative blast waves
(e.g. Sofue et al. \cite{sof80}; Kundu et al. \cite{kun80}). It
was further supported by the discovery of the associated pulsar
PSR J\,0538+2817 of comparable (characteristic) age (Anderson et
al. \cite{and96}).

The subsequent studies of the pulsar, however, lead to a
substantial reduction of the age of the system PSR
J\,0538+2817/SNR S\,147. The proper motion measurement of PSR
J\,0538+2817 combined with its angular offset from the geometric
center of S\,147 yields the kinematic age, $t_{\rm kin}$, of only
$\sim 3\times 10^4$ yr (Kramer et al. \cite{kra03}). Kramer et al.
(\cite{kra03}) pointed out that ``since SNRs typically fade away
after 100\,000 yr, pulsars genuinely associated with SNRs are
necessarily young". To reconcile the discrepancy between $t_{\rm
kin}$ and $\tau$, they suggested that the pulsar was born with a
spin period, $P_0 \simeq 0.139$ s, close to the present one (cf.
Romani \& Ng \cite{rom03})\footnote{Another possible way to reduce
the age of the pulsar is to assume that the pulsar's magnetic
field grows exponentially with a characteristic time-scale of
$\sim 4\tau /(n-\tilde{n} ) \, \simeq 2\times10^4$ yr, where
$n=3$, $\tilde{n} = \ddot{\nu} \nu /{\dot{\nu}} ^2$, and $\nu ,
\dot{\nu}$ and $\ddot{\nu}$ are, respectively, the pulsar spin
frequency and its two first time derivatives (measured by Kramer
et al. \cite{kra03}).} and therefore its true age is $\ll \tau$.
They also admit the possibility that S\,147 was formed by the SN
explosion in a low-density bubble blown-up by the SN progenitor's
wind (cf. Reich et al. \cite{rei03}), i.e. the Sedov-Taylor
solution cannot be used to estimate the parameters of the SN blast
wave, and particularly the age of the SNR (see Sect.\,3).

We agree that S\,147 could be the result of a cavity SN explosion
(see Sect.\,3) and note that in this case ({\it i}) the pulsar
birthplace could be significantly offset from the geometric center
of the SNR due to the proper motion of the SN progenitor star, and
therefore ({\it ii}) the proper motion vector should not
necessarily point away from this center (e.g. Gvaramadze
\cite{gva02}; Bock \& Gvaramadze \cite{boc02}; Gvaramadze
\cite{gva04}). Thus, ({\it i}) the true age of the system PSR
J\,0538+2817/SNR S\,147, $t_{\rm sys}$, could be either $\leq$ or
$\geq$ than $t_{\rm kin}$, and ({\it ii}) the association between
PSR J\,0538+2817 and S\,147 could be genuine even if more accurate
proper motion measurements will not prove that the pulsar is
moving away from the geometric center of the SNR.

Note also that the actual age of the system PSR J\,0538+2817/SNR
S\,147 is not fundamental for the further content of the paper
(see, however, Sect.\,5). The only thing we will assume in the
following is that $t_{\rm sys}$ (or the age of S\,147) is much
smaller than $\tau$ and that $\tau$ is nearly equal to the true
age of the pulsar. Even in the case of maximum possible offset of
the SN blast center from the geometric center of S\,147 and the
current position of PSR J\,0538+2817, $t_{\rm sys}\la 10^5$ yr.

Now we discuss whether or not the spectral characteristics of the
X-ray emission observed from PSR J\,0538+2817 are at variance with
our assumption that this pulsar is as old as indicated by its
characteristic age. Thus one should check whether or not the
effective temperature of the pulsar derived from model fits of its
X-ray spectrum agree with the theoretical expectations.

It is believed that the X-ray emission of middle-aged ($\sim 10^5$
yr) rotation-powered pulsars is predominantly of thermal origin
and consists of two components. The first one, the hard thermal
component, presumably originates from the hot polar caps heated
due to the pulsar activity, and the second one, the soft thermal
component, emerges from the rest of the surface of the pulsar. For
older ($\sim 10^6$ yr) pulsars, the stellar surface could be too
cold to be observable in X-rays, so that the X-ray emission of
these NSs comes mainly from the polar caps.

The blackbody fits to the X-ray spectra of PSR J\,0538+2817
obtained with the {\it Chandra X-Ray Observatory} (Romani \& Ng
\cite{rom03}) and the {\it XMM-Newton} (McGowan et al.
\cite{mcg03}) give the effective temperature $T_{\rm eff} \simeq
2.0-2.5\times 10^6$ K and the effective radius $R_{\rm eff} \simeq
1-3$ km of the spherical emitter. The inferred temperature is too
high to be consistent with the temperatures predicted by the
standard cooling models for NSs of age $\sim \tau$ (e.g. Yakovlev
\& Pethick \cite{yak04}), while small $R_{\rm eff}$ suggests that
the X-ray emission is associated with regions much smaller than
the entire surface of the NS. A possible interpretation of these
fits is that PSR J\,0538+2817 is a transition object between the
middle-aged pulsars and the older ones, and that the observed
X-ray emission is produced by the polar caps heated by the
bombardment of relativistic particles streaming down from the
pulsar acceleration zones (McGowan et al. \cite{mcg03}).

Alternatively, the above temperature inconsistency could be
considered as an indication that the surface of PSR J\,0538+2817
is not a perfect blackbody emitter and therefore its X-ray
spectrum should be treated in the framework of a NS atmosphere
model (with or without a magnetic field). McGowan et al.
(\cite{mcg03}) fitted the {\it XMM-Newton} spectrum of PSR
J\,0538+2817 with a nonmagnetic hydrogen atmosphere model. They
fixed the effective radius at values typical of NSs and found
$T_{\rm eff} \simeq 0.6-0.7\times 10^6$ K and $d\simeq 0.3-0.4$
kpc, where $d$ is the distance to the pulsar. Romani \& Ng
(\cite{rom03}) used the {\it Chandra} data to fit the spectrum of
the pulsar with a magnetic hydrogen atmosphere model and derived
about the same effective temperature, $T_{\rm eff} \simeq
0.65\times 10^6$ K; they also found $R_{\rm eff} =13$ km for $d$
fixed at 1.2 kpc (the figure based on the dispersion measure of
PSR J\,0538+2817; see Sect.\,2.4). Although the temperatures
derived in both models better agree with the standard cooling
curves, they are still somewhat higher than the predicted ones.
McGowan et al. (\cite{mcg03}) suggested that their fit should be
ruled out since it implies a much lower distance to the pulsar
than the generally accepted value of 1.2 kpc (see, however,
Sect.\,2.4).

Another possibility is that the pulsar PSR J\,0538+2817 belongs to
a class of very slowly cooling low-mass NSs with strong proton
superfluidity in their cores, whose cooling curves lie above the
basic standard cooling curve (see Yakovlev \& Pethick \cite{yak04}
and references therein).

Thus we conclude that the existing X-ray data do not contradict
the possibility that the true age of PSR J\,0538+2817 is $\simeq
\tau$.

\subsection{Distance to the system PSR J\,0538+2817/SNR S\,147}

Most of the existing distance estimates for S\,147 are based on
the highly unreliable empirical relationships between surface
brightness and linear diameter for SNRs. These estimates, ranging
from $d \simeq 0.7$ kpc (Milne \cite{mil70}) to $\simeq 1.6$ kpc
(Sofue et al. \cite{sof80}), cannot be restricted from H\,I
absorption measurements since S\,147 lies towards the Galactic
anticenter. A possible way to constrain the distance to S\,147
comes from the study of absorption lines in spectra of stars
located along the line-of-sight towards this extended SNR.
Numerous observations (e.g. Phillips et al. \cite{phi81}; Phillips
\& Gondhalekar \cite{phi83}; see also Sallmen \& Welsh
\cite{sal04}) have revealed the existence of high-velocity gas
(associated with the SNR's shell) towards two stars,
\object{HD\,36\,665} and \object{HD\,37\,318}, located,
respectively, at $\simeq 0.9$ and $\simeq 1.4$ kpc (these figures
were derived on the basis of spectral types, visual magnitudes and
color excesses of the stars). Thus the distance to S\,147 is $\la
0.9$ kpc, that is consistent with the distance of $\simeq 0.8$ kpc
derived by Fesen et al. (\cite{fes85}) from the interstellar
reddening to the SNR. Note that the latter two estimates should be
somewhat reduced if the reddening to S\,147 and the background
stars is enhanced by the dust associated with the SNR's shell (cf.
Gondhalekar \& Phillips \cite{gon80}). There are also several
stars in the direction of S\,147 with parallaxes measured with
{\it Hipparcos}. The most distant of them, \object{HD\,37\,367},
is located at $0.36 _{-0.09} ^{+0.15}$ kpc. The absorption
spectrum of this star does not show high-velocity lines (Sallmen
\& Welsh \cite{sal04}), that suggests that HD\,37\,367 is a
foreground star and puts a lower limit on the distance to S\,147
of $\sim 0.4$ kpc.

An indirect estimate of the distance to S\,147 comes from its
association with PSR J\,0538+2817. The dispersion measure of the
pulsar, DM$\simeq 40 \, {\rm pc} \, {\rm cm}^{-3}$ (Anderson et
al. 1996), and the Cordes \& Lazio (\cite{cor02})\footnote{See
also http://rsd-www.nrl.navy.mil/7213/lazio/ne\_model/} model for
the distribution of Galactic free electrons yield a distance to
the system of $\simeq 1.2\pm 0.2$ kpc, that is somewhat greater
than the upper limit on the distance to S\,147 of $\simeq 0.9$
kpc. Note that the Cordes \& Lazio model does not take into
account a possible contribution to the dispersion of the pulsar's
signal from the ionized material associated with the shell of
S\,147. If the excess dispersion measure due to the SNR's shell
$\Delta DM \ga 10 \, {\rm pc} \, {\rm cm}^{-3}$ (see Sect.\,3),
than the distance to the pulsar is $\la 0.9\pm 0.2$ kpc (Cordes \&
Lazio \cite{cor02}).

In the following we allow $d$ to vary in a wide interval from 0.4
to 0.9 kpc, and sometimes use the figure of 1.2 kpc (accepted in
the majority of recent papers devoted to PSR J\,0538+2817 and
S\,147) to compare our results with those of other authors. The
uncertainty in the distance, however, does not affect the main
results of the paper.

\section{S\,147 as the result of SN explosion within a WR bubble}

Let us now show that the small (kinematic) age of S\,147 and the
low expansion velocity of its shell cannot be reconciled with each
other if one assumes that the SN blast wave evolves in a
homogeneous, uniform medium, i.e. if one describes the evolution
of S\,147 in the framework of the standard Sedov-Taylor model or
models for radiative blast waves.

According to the Sedov-Taylor model, the expansion
velocity of the SN blast wave is given by $v_{\rm S-T} = 0.4
R_{\rm SNR}/t_{\rm SNR}$, where $R_{\rm SNR}$ and $t_{\rm SNR}$
are, respectively, the radius and the age of the SNR.
The use of this model implies that the SN blast center
coincides with the center of the SNR, and therefore $t_{\rm SNR}
=t_{\rm kin}$. For the angular radius of S\,147 of $\simeq 1\fdg
5$ and $t_{\rm SNR} \simeq 3\times 10^4$ yr, one has $v_{\rm S-T}
\simeq  340 \, d_1 \, {\rm km} \, {\rm s}^{-1}$, where $d_1$ is
the distance to the SNR in units of 1 kpc. This estimate agrees
with the observed expansion velocity $v_{\rm SNR} \simeq 80-120 \, {\rm km}
\, {\rm s}^{-1}$
if $d_1 = 0.24-0.35$, that is for distances smaller than the lower
limit given in Sect.\,2.4. Moreover, the Sedov-Taylor solution
implies the following estimate of the number density of the ambient
interstellar medium: $n_{\rm ISM} \simeq 0.24 \, d_{1} ^{-5} \,
{\rm cm}^{-3}$, i.e. $\simeq 300 \, {\rm cm}^{-3}$ for $d_1 = 0.24$
or $\simeq 50 \, {\rm cm}^{-3}$ for $d_1 = 0.35$. Both values of
$n_{\rm ISM}$ are inconsistent with the fundamental assumption
of the Sedov-Taylor model that the SN blast wave is adiabatic.

The use of radiative models (e.g. Cioffi et al. \cite{cio88};
Blondin et al. \cite{blo98}) does not improve the situation. For
example, using the relationships given in Cioffi et al.
(\cite{cio88}; see their Eqs.\,(3.32-3.33)), one has that $v_{\rm
SNR}$ could be consistent with the theoretical value if $n_{\rm
ISM} \simeq 10 \, {\rm cm}^{-3}$ (for $v_{\rm SNR} \simeq 120 \,
{\rm km} \, {\rm s}^{-1}$) or $n_{\rm ISM} \simeq 30 \, {\rm
cm}^{-3}$ (for $v_{\rm SNR} \simeq 80 \, {\rm km} \, {\rm
s}^{-1}$). In turn, these density estimates imply the distance to
the SNR of, respectively, $\simeq 0.41$ and $\simeq 0.32$ kpc.

An alternative to these models is the possibility that S\,147 is
the remnant of a SN which exploded within a low-density bubble
surrounded by a shell (created by the stellar wind of the SN
progenitor). In this case, the expansion velocity of the SN blast
wave could be small even for young SNRs (see below). We believe
that S\,147 is the result of a SN explosion within the bubble
blown-up during the WR phase of evolution of the SN progenitor
star. Our belief is based on the interpretation of the general
structure of S\,147, whose shell is characterized by bilateral
symmetry and is elongated along the symmetry axis.

The bilateral and elongated appearance of some SNRs implies that
the regular interstellar magnetic field is involved in shaping
their shells. However, it was recognized long ago that the tension
associated with the interstellar magnetic field cannot directly
affect the shape of a typical SN blast wave to cause it to be
elongated (e.g. Manchester \cite{man87}). On the other hand, the
regular magnetic field could affect the symmetry of large-scale
structures created in the interstellar medium by virtue of the
ionizing emission and stellar wind of massive stars -- the
progenitors of most of SNe. For example, the elongated bilateral
SNRs could originate if the SN blast waves in these SNRs take on
the shape of the wind bubbles blown-up by the SN progenitor stars
during the main-sequence phase and distorted by the surrounding
regular magnetic field (Arnal \cite{arn92}; Gaensler
\cite{gae98}). But the relatively small size of S\,147 argues
against a SN explosion within the main-sequence bubble and
suggests that the pre-existing bubble was rather blown-up during
the (much shorter) WR phase. Additional support for this
suggestion comes from the low expansion velocity of the SNR's
shell, which could be naturally explained if the wind bubble was
surrounded by a massive shell -- the distinctive feature of WR
bubbles (see Gvaramadze \cite{gva04}). In the presence of the
regular interstellar magnetic field the structure of the WR shell
is modified in such a way that the density distribution over the
shell acquires an axial symmetry with the minimum column density
at the magnetic poles (see Gvaramadze \cite{gva04} and references
therein). A good example of a bilateral WR shell is the nebula
\object{S\,308} around the WR star \object{HD\,50896} (see, e.g.,
Fig.\,1c of van Buren \& McCray \cite{van88}). The subsequent
interaction of the SN blast wave with the magnetized WR shell
results in the origin of a bilateral SNR with two blow-ups along
its symmetry axis.

The low expansion velocity of the SNR's shell could be treated as
an indication that the pre-existing (WR) shell was massive enough.
Numerical simulations by Tenorio-Tagle et al. (\cite{ten91})
showed that the SN blast wave merges with the wind-driven shell
and the resulting SNR enters into the momentum-conserving stage
(i.e. $v_{\rm SNR} \sim 100 \, {\rm km} \, {\rm s}^{-1}$) if the
mass of the shell is $\ga 50\, M_{\rm ej}$, where $M_{\rm ej}$ is
the mass of the SN ejecta. Assuming that the radius of the WR
shell $R_{\rm WR}\simeq 20$ pc ($d=0.9$ kpc) and $M_{\rm ej}
\simeq 4 M_{\odot}$ (see Sect.\,4), one has the number density of
the ambient interstellar medium $n_{\rm ISM} \geq 0.2 \, {\rm
cm}^{-3}$. This estimate could be further constrained if one uses
the result by Tenorio-Tagle et al. (\cite{ten91}) that the
reaccelerated wind-driven shell (now the shell of the SNR)
acquires $\sim 10\%$ of the initial SN energy $E_0 =10^{51}$ erg,
i.e. $v_{\rm SNR} \simeq (0.2 E_0 /M_{\rm shell})^{1/2}$, where
$M_{\rm shell} = (4\pi /3) R_{\rm WR} ^3 \rho _{\rm ISM}$, $\rho
_{\rm ISM} =1.4m_{\rm H} n_{\rm ISM}$, and $m_{\rm H}$ is the mass
of the hydrogen atom. For $v_{\rm SNR} =100\, {\rm km} \, {\rm
s}^{-1}$, one has $n_{\rm ISM} \simeq 1.0 \, {\rm cm}^{-3}$. The
latter figure could be used to estimate the excess dispersion
measure caused by the SNR's shell, $\Delta DM \simeq R_{\rm SNR}
n_{\rm ISM} /3 \simeq 8\, (d/0.9 \, {\rm kpc})^{-2} \, {\rm pc} \,
{\rm cm}^{-3}$, where $R_{\rm SNR} \simeq 23 \, (d/0.9 \, {\rm
kpc})$ pc is the characteristic radius of the SNR (cf.
Sect.\,2.4). This excess could be much larger if a large-scale
deformation of the SNR's shell (caused by the development of the
Richtmaier-Meshkov and Rayleigh-Taylor instabilities in the
reaccelerated wind-driven shell; see Gvaramadze \cite{gva99})
increases the line-of-sight extent of the ionized gas against the
pulsar (cf. Gvaramadze \cite{gva01}).

The north-south asymmetry of S\,147 could be understood if the WR
shell was swept up from the medium with the density growing to the
south (i.e. perpendicular to the orientation of the local regular
magnetic field, implied by the bilateral symmetry of the SNR's
shell). In this case, the SN blast wave merges with the south
(more massive) half of the WR shell and this part of the SNR
repeats the circular shape of the pre-existing shell. In the north
direction the blast wave merges only with the equatorial part of
the WR shell (where the column density is enhanced due to the
magnetic effect mentioned above) and overruns the less massive
segments of the shell in the northeast and northwest directions,
thereby producing two lobes in the north half of S\,147. Our
interpretation of the north-south asymmetry is testable. It would
be interesting to measure (e.g. with a Fabry-P\'{e}rot
interferometer; see Lozinskaya \cite{loz76}) the expansion
velocity of the northeast and northwest segments of S\,147 to
check whether or not it exceeds the characteristic expansion
velocity of the SNR's shell of $\sim 100 \, {\rm km} \, {\rm
s}^{-1}$.

Note that the above arguments in favour of cavity SN explosion
should be relevant independent of whether S147 is the result of
the SN explosion of a single (WR) star or of the (second) SN
explosion in a massive binary system.

\section{PSR J\,0538+2817 as the remnant of the first SN explosion in a
massive binary}

\subsection{Constraints on the parameters of the binary system}

In Sect.\,3 we suggested that the progenitor of the SN that
created S\,147 was a WR star, i.e. a massive star with the
zero-age main-sequence (ZAMS) mass $\ga 20\, M_{\odot}$ (e.g.
Vanbeveren et al. \cite{van98}). Let us assume that this SN
explosion was the second one in a massive binary and that the
pulsar PSR J\,0538+2817 is the remnant of the first SN explosion.
The latter assumption implies that the ZAMS mass of the first SN
progenitor was $\leq 25-30 \, M_{\odot}$ (more massive progenitors
produce black holes). We assume also that PSR J\,0538+2817 is as
old as indicated by its characteristic age, i.e. the SN explosions
were separated by a time scale of $\sim \tau$. From this it
follows that the ZAMS masses of the binary components were nearly
equal to each other (cf. Bethe \& Brown \cite{bet98}; Vlemmings et
al. \cite{vle04}), so that it is likely that the second SN
explosion also forms a NS.

The spin characteristics ($P$ and $\dot{P}$) and the (inferred)
magnetic field of PSR J\,0538+2817 are typical of non-recycled
pulsars. One can conclude therefore that the binary system was
sufficiently wide so that the stellar wind of the massive
companion star did not appreciably affect the evolution of the
pulsar, i.e. the standoff radius of the pulsar wind, $r_{\rm s}$,
was larger than the accretion radius, $r_{\rm a}$ (cf. Illarionov
\& Sunyaev \cite{ill85}):
\begin{equation}
r_{\rm s} (t) \equiv (|\dot{E}(t)| /4\pi \rho _{\rm w} c v_{\rm w} ^2
)^{1/2} \, > r_{\rm a} \equiv 2GM_{\rm p}/v_{\rm w} ^2 \,  ,
\label{separ}
\end{equation}
where $c$ is the speed of light, $\rho _{\rm w} =\dot{M} _{\rm w}
/4\pi a^2 v_{\rm w}$, $\dot{M} _{\rm w}$ and $v_{\rm w}$ are,
respectively, the mass-loss rate and wind velocity of the
companion star, $a$ is the binary separation, $G$ is the
gravitational constant, and $M_{\rm p} =1.4 \, M_{\odot}$ is the
mass of the pulsar. From Eq.\ (\ref{separ}) one has that $a$
should be larger than some critical value given by the following
relationship:
\begin{equation}
a_{\rm cr} \,  = \, {2GM_{\rm p} \over v_{\rm w} ^2} \left(
{\dot{M} _{\rm w} v_{\rm w} c \over |\dot{E_{\star}}|} \right)
^{1/2} \, , \label{crit}
\end{equation}
where $|\dot{E_{\star}}| =32\pi ^4 \mu ^2 /3c^3 P_{\star} ^4$ and
$P_{\star} = (P^2 - 16\pi ^2 \mu ^2 t_{\rm sys} /3c^3 I)^{1/2}$
are the spin-down luminosity and the spin period of the pulsar at
the moment of the second SN explosion, $\mu = BR^3$ is the
magnetic moment of the pulsar (we assume that $\mu = {\rm
const}$), $R=10$ km is the radius of the pulsar, and $t_{\rm sys}
\simeq 3\times 10^4$ yr.

One can envisage two possible situations at the moment of the
first SN explosion. First, the companion star was a red supergiant
and its extended envelope survived the passage of the SN blast
wave. Second, the convective envelope of the companion red
supergiant star was blown up by the SN ejecta to leave a bare He
core (i.e. a WR star) or the companion star had already entered
into the WR phase. In the first situation, $\dot{M}_{\rm w} \equiv
\dot{M}_{\rm w} ^{\rm RSG} \simeq 10^{-5} \, M_{\odot} \, {\rm
yr}^{-1}$ and $v_{\rm w} \equiv v_{\rm w} ^{\rm RSG} \simeq 10 \,
{\rm km} \, {\rm s}^{-1}$, so that one has from Eq.\,(\ref{crit})
that $a_{\rm cr} \simeq 10^5 \, R_{\odot}$. In such a wide system
the stripping and ablation of the red supergiant envelope can be
neglected. In the second situation, $\dot{M} _{\rm w} \equiv
\dot{M} _{\rm w} ^{\rm WR} \simeq 10^{-5} \, M_{\odot} \, {\rm
yr}^{-1}$ and $v_{\rm w} \equiv v_{\rm w} ^{\rm WR} \simeq 2\,000
\, {\rm km} \, {\rm s}^{-1}$, so that one has $a_{\rm cr} \simeq
35\, R_{\odot}$. Thus if $a\geq a_{\rm cr}$, the pulsar was active
during the whole period between the two SN explosions and its spin
characteristics were not appreciably affected by the wind of the
companion star; the latter implies that the true age of the pulsar
is equal to $\tau$ (provided that $P_0 << P$).

\subsection{Origin of the pulsar peculiar velocity}

Proper motion measurement of PSR J\,0538+2817 by Kramer et al.
(\cite{kra03}) yields a pulsar transverse velocity of $v_{{\rm
p},{\lambda}} = 194\pm 14 \, d_1  \, {\rm km} \, {\rm s}^{-1}$, if
one uses the proper motion in the ecliptic longitude only, or
$v_{\rm p} = 318_{-100} ^{+230} \, d_1 \, {\rm km} \, {\rm
s}^{-1}$ for the composite proper motion. For $d_1 =0.4-0.9$, one
has $v_{{\rm p},{\lambda}} \simeq 80-170 \, {\rm km} \, {\rm
s}^{-1}$ and $v_{\rm p} \simeq 130-290 \, {\rm km} \, {\rm
s}^{-1}$. Using the estimates of $a_{\rm cr}$ derived in
Sect.\,4.1, one can check whether or not $v_{{\rm p},{\lambda}}$
and $v_{\rm p}$ are consistent with the velocity, $v_{\rm NS}
^{\rm old}$, of the old NS released from orbit by the second SN
explosion.

In the case of a symmetric SN explosion and a circular binary
orbit one has
\begin{equation}
v_{\rm NS}  ^{\rm old} \, = \,  \left( {m^2 - 2m -2 \over m+1}
\right)^{1/2} \, \left({GM_{\rm p} \over a} \right)^{1/2} \, ,
\label{sym}
\end{equation}
where $m=M/M_{\rm p}$ and $M\la 5-6 M_{\odot}$ (e.g. Vanbeveren et
al. \cite{van98}) is the mass of the pre-SN star. It is clear that
$v_{\rm NS} ^{\rm old}$ is inconsistent with both $v_{{\rm
p},{\lambda}}$ and $v_{\rm p}$ if the second SN exploded after the
red supergiant phase (i.e. $a\geq 10^5 \, R_{\odot}$). In the
second situation (i.e. $a \geq 35 \, R_{\odot}$), one has from
Eq.\,(\ref{sym}) that $v_{\rm NS}  ^{\rm old} \leq 80-110 \, {\rm
km} \, {\rm s}^{-1}$. This estimate shows that $v_{\rm NS} ^{\rm
old}$ is inconsistent with $v_{\rm p}$, but could be equal to
$v_{{\rm p},{\lambda}}$ if one adopts the smallest values of $a$
and $d_1$ and the largest one of $M$, and provided that the radial
component of the pulsar velocity is small.

One can, however, assume that the SN explosion was asymmetric so
that the young NS received a kick velocity, $w$, at birth. In this
case, the new-born NS can impart some momentum to the old NS in
the course of disintegration of the binary system (see Tauris \&
Takens \cite{tau98}). The magnitude of the momentum depends on the
angle, $\theta$, between the kick vector and the direction of
motion of the exploding star, and the angle, $\phi$, between the
kick vector and the orbital plane (see Fig.\,1 of Tauris \& Takens
\cite{tau98})\footnote{For $\phi =0^{\degr}$ and $0^{\degr} \leq
\theta \leq 180^{\degr}$, the kick vector lies in the orbital
plane and points to the half-sphere occupied by the old NS.}. An
analysis of Eqs.\,(44)-(47) and (51)-(56) given in Tauris \&
Takens (\cite{tau98}) shows that the momentum imparted to the old
NS is maximum if
\begin{equation}
\theta \, \sim \, \theta _{\ast}  \, \equiv \, \arccos(-v/w) \, ,
\label{theta}
\end{equation}
where $v=[G(M+M_{\rm p} )/a]^{1/2}$ is the relative orbital
velocity, and provided that the vector of the kick velocity does
not strongly deviate from the orbital plane of the binary system;
i.e. than the kick received by the second-born NS is directed
almost towards the old NS.
\begin{figure}
 \resizebox{8cm}{!}{\includegraphics{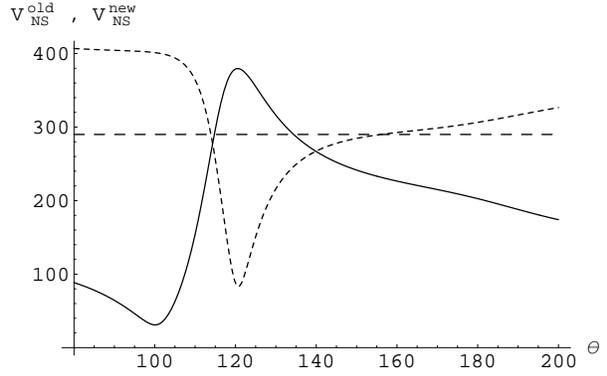}}
 \caption{The dependence of the velocities of the old and new-born
neutron stars (shown, respectively, by the solid and the
short-dashed lines) on the angle between the kick vector and the
direction of motion of the exploding star. The long-dashed line
shows the best-fit transverse velocity at the upper distance 0.9
kpc. See text for details.}
  \label{twovel}
\end{figure}
Fig.~\ref{twovel} illustrates how the direction of the kick
affects the velocities of the old and new-born NSs released from
the disrupted binary. One can see that for $\theta \simeq \theta
_{\ast} = 118^{\degr}$ (we assume that $w=400 \, {\rm km} \, {\rm
s}^{-1} , M=5M_{\odot}$ and $a=35 \, R_{\odot}$, and that $\phi
=0^{\degr}$) $v_{\rm NS} ^{\rm old}$ is maximum ($\sim w$), while
$v_{\rm NS} ^{\rm new}$ drops to a minimum value of $\simeq 0.25
w$. It is also seen that for $\theta$ ranging from $\simeq
115^{\degr}$ to $\simeq 135^{\degr}$ $v_{\rm NS} ^{\rm old}$ is
larger than the best-fit transverse velocity at the upper distance
0.9 kpc. Note that the maximum value of $v_{\rm NS} ^{\rm old}$ is
almost independent of $a$ (or $v$): $v_{\rm NS} ^{{\rm old}, {\rm
max}}$ tends to $w$ with increasing $a$\footnote{Vlemmings et al.
(\cite{vle04}) draw an erroneous conclusion that in a wide binary
disrupted after the second (asymmetric) SN explosion the old NS
could be accelerated to a velocity $\sim w$, while the second-born
NS to a velocity several times exceeding $w$.}, while $v_{\rm NS}
^{\rm new}$ drops to 0.
\begin{figure}
 \resizebox{8cm}{!}{\includegraphics{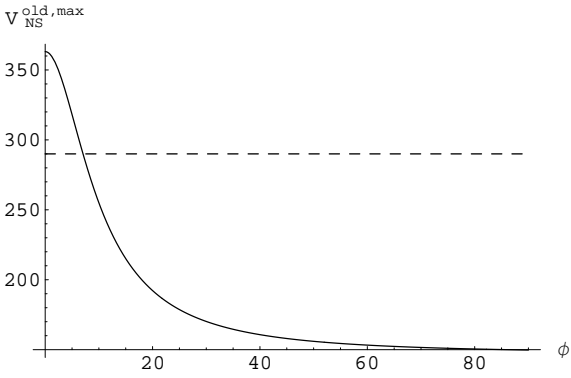}}
 \caption{The dependence of the maximum value of the velocity of
 the old neutron star on the angle between the kick vector and
 the orbital plane.}
  \label{velphi}
\end{figure}
Fig.~\ref{velphi} shows how $v_{\rm NS} ^{{\rm old},{\rm max}}$
depends on $\phi$; it is seen that $v_{\rm NS} ^{{\rm old},{\rm
max}} \ga v_{\rm p} ^{\rm max}$ if $\phi \la 10^{\degr}$. The
above considerations show that $v_{\rm NS}^{\rm old}$ could be
consistent with $v_{\rm p}$, but the binary separation must be as
small as possible while avoiding pulsar recycling and the kick
direction must be carefully tuned. For $d_1 =0.9$ and $a \sim 35
R_{\odot}$ the probability of a favourable kick orientation is
$\sim 10^{-3}$ (for the isotropic kick distribution) or $\sim
10^{-2}$ (if kicks produced by SN explosions in binary systems are
restricted close to the orbital plane; see Sect.\,5). Note also
that the smaller $d_1$ the wider the range of angles for which
$v_{\rm NS} ^{\rm old} > v_{\rm p}$ and the larger the probability
that the second-born NS will receive an appropriately oriented
kick.

\section{Discussion}

In Sect.\,4.2 we considered two possible explanations for the
origin of the peculiar velocity of PSR J\,0538+2817 based on the
idea that the pulsar could be the remnant of the first SN
explosion in a massive binary. Proceeding from this we found that
although the pulsar motion in the ecliptic longitude taken
separately (cf. Hobbs et al. \cite{hob05}; Lewandowski et al.
\cite{lew04}) could be consistent with the possibility that the
binary was disrupted by a symmetric SN explosion, the large
composite proper motion measured by Kramer et al. (\cite{kra04};
if confirmed by more accurate measurements) makes the case of an
asymmetric SN explosion more plausible. In this case, we expect
that the kick velocity received by the second-born NS is
restricted close to the orbital plane (i.e. $\phi \la 10^{\degr}$;
cf. Wex et al. \cite{wex00}).
\begin{figure}
 \resizebox{8cm}{!}{\includegraphics{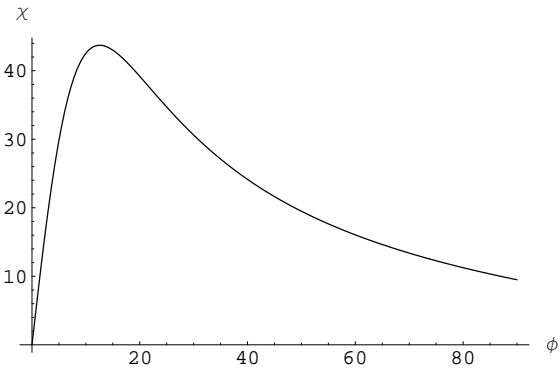}}
 \caption{The angle between the velocity vector
of the old neutron star and the orbital plane as a function of the
angle between the kick vector and the orbital plane.}
  \label{chiphi}
\end{figure}
Fig.~\ref{chiphi} shows the angle, $\chi$, between $v_{\rm NS}
^{\rm old}$ [for $\theta$ given by Eq.\,(\ref{theta})] and the
pre-SN orbital plane as a function of $\phi$: $\chi$ grows from
$0^{\degr}$ (for $\phi =0^{\degr}$) to $\simeq 40^{\degr}$ (for
$\phi \simeq 10^{\degr}$) and than gradually decrease to $\simeq
10^{\degr}$ for $\phi =90^{\degr}$. Thus if our explanation of the
age discrepancy is correct, we do not expect any alignment between
the pulsar spin axis and proper motion, although one cannot
exclude that they are perpendicular to each other if the second SN
explosion was symmetric about the orbital plane (i.e. $\phi \simeq
0^{\degr}$; cf. Cordes \& Wasserman \cite{cor84}; Colpi \&
Wasserman \cite{col02}; Vlemmings et al. \cite{vle04}) and
provided that the spin axis of PSR J\,0538+2817 was perpendicular
to this plane. To check whether this possibility was realized one
must know the precise position angles of the pulsar proper motion
and the symmetry axis of the X-ray nebula surrounding the pulsar.
However, the large error in latitudinal direction makes the
position angle of the pulsar proper motion very uncertain (see
Sect.\,2.2), while the low photon statistics of the {\it Chandra}
data does not allow us to infer the symmetry axis of the pulsar
wind nebula unambiguously. We will discuss this point in more
detail.

The existing {\it Chandra} data on PSR J\,0538+2817 show that the
nebula around the pulsar consists of a core and a dim halo (Romani
\& Ng \cite{rom03}). These substructures could be fitted with
ellipses with the semi-major axes of $2\farcs5$ and $9^{\arcsec}$,
at the position angles of $\sim 130^{\degr}$ and $\sim
60^{\degr}$. Romani \& Ng (\cite{rom03}) suggested that the halo
``has a scale comparable to the expected wind shock radius for a
confinement pressure appropriate to the interior of S\,147" and
used the geometry of the halo to infer the pulsar spin axis at a
position angle of $334\fdg0 \pm 5\fdg5$ (that is consistent with
the position angle of the pulsar composite proper motion). Their
calculations of the shock radius were carried out in the framework
of the Sedov-Taylor model and under the assumption of a
spherically-symmetric pulsar wind.

The angular radius of the toroidal shock can be expressed as
\begin{equation}
\Theta \, \simeq \, (|\dot{E} | /4\pi \alpha c d^2 P_{\rm in}
)^{1/2} \, , \label{tor}
\end{equation}
where $\alpha \leq 1$ is a factor characterizing the anisotropy of
the energy flux of the pulsar wind (for $\alpha <<1$ most of the
pulsar wind is confined to the equatorial plane, while $\alpha =1$
corresponds to the spherically-symmetric wind),
\begin{equation}
P_{\rm in} ={(\gamma -1)E_{\rm th} \over V} \, ,
\label{pres}
\end{equation}
is the mean gas pressure inside the SNR, $\gamma$ is the specific
heat ratio, $E_{\rm th} =\delta E_0$ is the thermal energy of the
SN blast wave, $\delta <1$, and $V=(4\pi/3)R_{\rm SNR} ^3$ is the
volume of the SNR. For adiabatic gas ($\gamma =5/3)$ and an
angular radius of the SNR of $\sim 1\fdg5$, one has from
Eqs.\,(\ref{tor}) and (\ref{pres}) that
\begin{equation}
\Theta \, \simeq \, 1\farcs4 \, (\alpha \delta) ^{-1/2} \, d_1
^{1/2} \, . \label{torus}
\end{equation}

For a Sedov-Taylor blast wave (i.e. $\delta \simeq 0.72$) and
assuming $\alpha =1$ and $d_1 =1.2$ (Romani \& Ng \cite{rom03}),
one has $\Theta =1\farcs8$, that is consistent with the angular
radius of the core of the nebula rather than with the scale of the
more extended halo. Moreover, in Sect.\,3 we showed that the
Sedov-Taylor model cannot reconcile the small expansion velocity
of S\,147 with its low kinematic age and suggested that this SNR
is the result of a SN explosion within a pre-existing bubble
surrounded by a massive wind-driven shell. In this case, the
evolution of the SN blast wave completely differs from that of the
Sedov-Taylor one. The presence of the massive shell strongly
increases the radiative cooling of the SN blast wave, that results
in a rapid decrease of the thermal content of the SNR as compared
with the Sedov-Taylor one. Numerical simulations by Tenorio-Tagle
et al. (\cite{ten91}; see their Fig.\,3) show that $\delta \simeq
0.4$ at $t_{\rm SNR} =3\times 10^4$ yr and further decreases to
$\simeq 0.1$ at $t_{\rm SNR} = 10^5$ yr. Thus assuming that
$t_{\rm SNR} = t_{\rm kin} = 3\times 10^4$ yr, one has from
Eq.\,(\ref{torus}) that $\Theta \leq 2\farcs1$ if the distance to
S\,147 is $\leq 0.9$ kpc, or $\simeq 2\farcs4$ if $d_1=1.2$. These
estimates also suggest that the toroidal shock could be associated
with the core of the pulsar wind nebula. However, in the case of a
cavity SN explosion $t_{\rm kin}$ could be as large as $\sim 10^5$
yr (see Sect.\,2.3); in this case $\delta \sim 0.1$ and $\Theta
\simeq 3\farcs6 -4\farcs7$ (for $d_1$ ranging from 0.9 to 1.2),
that better agrees with the scale of the halo (cf. Romani \& Ng
\cite{rom03}). The agreement could also be achieved when the
anisotropy of the pulsar wind is taken into account. Assuming that
$\alpha =0.1$, one has from Eq.\,(\ref{torus}) that $\Theta \leq
6\farcs6$ and $7\farcs6$ for $d_1 = 0.9$ and 1.2, respectively.

If the core indeed corresponds to the equatorial torus of the
pulsar wind nebula (this could be checked with forthcoming deep
{\it Chandra} observations\footnote{See
http://heasarc.gsfc.nasa.gov/cgi-bin/W3Browse/w3table.pl} or
high-resolution radio imaging), than the position angle of the
symmetry axis of the torus is exactly perpendicular to the
direction of the pulsar composite proper motion (interferometric
measurements of PSR J\,0538+2817 would be highly valuable for
further restricting the position angle of the pulsar proper
motion), i.e. just what is expected if the pulsar obtained its
peculiar velocity by the disintegration of the binary system (with
aligned angular momenta) by symmetric SN explosion or by SN
explosion symmetric about the orbital plane (see above). Note,
however, that the same spin-kick orientation could be produced by
the SN explosion of a solitary massive star if the spin and the
peculiar velocity of the pulsar are due to a single off-center
kick (Spruit \& Phinney \cite{spr98}).

Now we discuss the problem of the second (young) NS possibly
associated with S\,147. The detection of such a NS would allow us
not only to distinguish the case of a SN explosion in a binary
system from that of a solitary SN explosion, but also would lend
strong support to our explanation of the age discrepancy and would
have a strong impact on our understanding of the kick physics.

In the case of a symmetric SN explosion the angle, $\psi$, between
the velocity vectors of the old and new-born NSs depends only on
the mass of the SN ejecta and is given by (Gott et al.
\cite{got70}; see also Iben \& Tutukov \cite{ibe96})
\begin{equation}
\psi \, = \, \arccos \left[4\left({M_{\rm ej} \over 2M_{\rm p}
}\right)^2 -3\right]^{-1/2} \, , \label{psi}
\end{equation}
where $M_{\rm ej} =M-M_{\rm p}$. It follows from Eq.\,(\ref{psi})
that $\psi$ is always smaller than $90^{\degr}$; e.g. $\psi \sim
60^{\degr}$ for the parameters adopted above. In the case of an
asymmetric SN explosion $\psi$ could be larger than $90^{\degr}$.
\begin{figure}
 \resizebox{8cm}{!}{\includegraphics{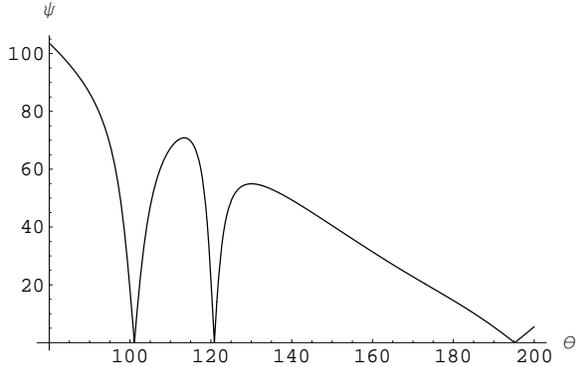}}
 \caption{The angle between the velocity vectors of the old and
new-born neutron stars as a function of the angle between the kick
vector and the direction of motion of the exploding star.}
  \label{chitheta}
\end{figure}
Fig.~\ref{chitheta} shows how $\psi$ depends on the direction of
the kick vector (we assume here that $\phi =0^{\degr}$). It is
seen that $\psi$ is very sensitive to $\theta$. For $\theta$
ranging from $115^{\degr}$ to $135^{\degr}$ (i.e. when $v_{\rm NS}
^{\rm old} > v_{\rm p} ^{\rm max}$) $\psi =0^{\degr}$ for $\theta
\sim \theta _{\ast}$ [see Eq.\,(\ref{theta})] and grows to $\simeq
60^{\degr} -70^{\degr}$ at the bounds of the range. Thus we expect
that the young NS should be located in a cone with a half-opening
angle $\la 70^{\degr}$ with the cone axis oriented along the
proper motion of PSR J\,0538+2817. We also expect that the young
NS should be situated closer to the vertex of the cone (i.e. to
the SN blast center) than PSR J\,0538+2817 since $v_{\rm NS} ^{\rm
new} < v_{\rm NS} ^{\rm old}$ (the same is true for the case of a
symmetric SN explosion). Remind that the SN blast center could be
significantly offset from the geometric center of S\,147 (see
Sect.\,2.3).

The most optimistic supposition is that the young NS is an
ordinary (rotation-powered) pulsar with a favourably oriented
radio beam. The youth of the pulsar implies that it should be more
energetic than PSR J\,0538+2817, and therefore could be easily
detected somewhere to the south of the $1^{\degr}$ strip covered
by the Arecibo survey (most likely in the west half of the SNR).
Another possibility is that the young NS is an off-beam radio
pulsar or that it belongs to a class of radio-quiet NSs. In both
cases one can expect that the young NS should be a sufficiently
bright soft X-ray source to be detected with the {\it ROSAT}
All-Sky Survey (RASS). Indeed, the RASS Faint Source
Catalog\footnote{http://www.xray.mpe.mpg.de/rosat/survey/rass-fsc/}
shows five sources in the proper place. The count rates of these
sources (ranging from $\simeq 0.014$ to $\simeq 0.039 \, {\rm ct}
\, {\rm s}^{-1}$) should be compared with the count rate in the
{\it ROSAT} pass band expected for a NS of age of $\simeq 3\times
10^4$ yr. Let us assume that the new-born NS is a spherical
blackbody emitter of radius of 10 km and temperature of $\simeq
10^6$ K (predicted by standard cooling models). Then assuming an
interstellar absorption column density of $\simeq 2.5-3\times
10^{21} \, {\rm cm}^{-2}$ (e.g. McGowan et al. \cite{mcg03};
Romani \& Ng \cite{rom03}) and using
PIMMS\footnote{http://heasarc.gsfc.nasa.gov/Tools/w3pimms.html},
one has $\simeq 0.017 - 0.022$ {\it ROSAT} PSPC ${\rm ct} \, {\rm
s}^{-1}$, that is, the figures comparable with the above count
rates. Thus one cannot exclude that one of the five RASS sources
is a young NS associated with S\,147. If this NS is a
rotation-powered pulsar, its position should be marked by a pulsar
wind nebula with a characteristic scale of at least an order of
magnitude larger than that of the nebula around PSR J\,0538+2817.
Deep radio or X-ray observations would allow us to verify the
existence of such a nebula.

Note that the RASS sources are about 1.5-4 times fainter than the
pulsar PSR J\,0538+2817 (also detected by the RASS; e.g. Sun et
al. \cite{sun96}), while the opposite situation is expected if PSR
J\,0538+2817 is as old as indicated by its spin-down age and if
its cooling follows the standard cooling curves. The
contradiction, however, could be removed if (as discussed in
Sect.\,2.3) the pulsar belongs to a class of slowly cooling NSs.

One cannot exclude that the young stellar remnant collapsed into a
black hole. In this case, the chances of finding this object are
small.

\begin{acknowledgements}
I am grateful to A.M.Cherepashchuk, A.V.Tutukov and A.A.Vikhlinin
for useful discussions, to the IPHAS collaboration and personally
to A.A.Zijlstra for providing the electronic version of the ${\rm
H}_{\alpha}$ image of S\,147, and to the anonymous referee for
useful suggestions.
\end{acknowledgements}

\end{document}